\documentclass{article}
\usepackage{spconf,amsmath,graphicx,hyperref}
\usepackage{booktabs}
\usepackage{multirow}
\usepackage{svg}
\usepackage{enumitem}

\title{ASR ensembling for phoneme intelligibility evaluation of speech anonymizers}

\name{Victor Ménestrel$^{\dagger\ddagger}$, Sebastian Möller$^{\dagger}$, Slim Ouni$^{\ddagger}$, Dorothea Kolossa$^{\dagger}$}

\address{$^{\dagger}$Technische Universität Berlin, Faculty IV, Berlin, Germany\\
         $^{\ddagger}$Université de Lorraine, CNRS, Inria, Loria, F-54000, Nancy, France\\
         }

\begin{document}

\maketitle

\begin{abstract}
We present the first phoneme-level intelligibility evaluation of speech
anonymizers, assessing the performance of ASR-ensemble-based
metrics against measured intelligibility from a crowdsourced listening test. Our results show that simple hard-voting ASR metric reaches
correlations above $0.9$ with human ratings when aggregated by feature, test-type, or condition, provided that multiple ASR
models are combined; evaluating stimuli with and without a carrier sentence further improves the correlation at the stimulus level. However, posterior-probability-based confidence metrics bring no gain, which can be traced back to the insufficient calibration of the state-of-the-art open ASR models that were utilized here.
All data, code, and evaluation tools
are released as open source.
\end{abstract}


\section{Introduction}
\label{sec:intro}

Beyond the linguistic content, speech inherently encodes a wide range of biometric and personal attributes. Voice anonymization aims to suppress or modify the attributes that may leak a speaker's identity while preserving as much of the signal's utility as possible. Among the utility criteria, content preservation, also referred to as intelligibility, is the predominant one, as the anonymized speech must remain fully understandable to the listener.

Intelligibility is rarely measured by listening tests in the voice anonymization literature, even though this is the gold standard for assessing intelligibility \cite{chiang2023study}. Meyer et al.~\cite{meyer2025use}, in their analysis of use cases across 110 articles related to voice anonymization, report that less than 4\% conduct such actual intelligibility tests. Instead, 72\% use Automatic Speech Recognition (ASR) to measure intelligibility, mostly following the Voice Privacy Challenge (VPC) framework consisting in computing a Word Error Rate (WER) on the LibriSpeech test set \cite{miaovoiceprivacyb}. This WER depends heavily on the evaluation corpus: proper nouns and homophones can produce transcription errors even when the speech is perfectly intelligible to a human listener \cite{menestrellimitations}. The stimuli consist of meaningful sentences, which introduces a semantic-context bias, where ASR models can guess masked content from the linguistic context. Human listening tests, by contrast, can offer fine-grained perceptual insights but are costly and time-consuming to conduct. Natural speech also comprises sub-word phenomena, such as fillers and hesitations, which carry meaningful information for both privacy and utility \cite{franzrebsalgado2026evaluating}. Thus, phoneme-level evaluation could prove valuable for analyzing the preservation of spoken content. This situation gives rise to a paradox: human intelligibility is widely recognized as important, yet researchers increasingly rely on automated biased metrics. Indeed, Panariello et al.~\cite{panariello2024voiceprivacya} showed a $0.14$ Pearson correlation between $1-WER$ and perceived intelligibility measured by listening tests on speech anonymizers. Crucially, prior listening tests have mostly assessed perceived rather than actual intelligibility. Perceived intelligibility asks listeners to rate, typically on a Likert scale, how intelligible they judge the speech to be. Actual intelligibility measures whether listeners can truly understand the content, by transcribing or reporting what they heard.

In this work, we propose a crowdsourcing-based approach to evaluate the phoneme-level intelligibility of speech anonymizers. We introduce metrics that approximate human judgments by aggregating the outputs of an ensemble of ASR models, yielding a fine-grained diagnosis of intelligibility. These metrics distinguish word-initial from word-final confusions and break down performance by six distinctive features. They achieve a noticeable correlation with human ratings, allowing a fast and low-cost intelligibility diagnosis.

All code and the anonymized collected data are released as open source, with the goal of allowing any researcher to evaluate the performance of their anonymizer \footnote{https://github.com/VictorMenestrel/DALT-for-Speech-Anonymizers}.

\section{Methodology}
\label{sec:pagestyle}

\subsection{Speech Corpus}
\label{ssec:subhead1}

Our evaluation is based on the English Diagnostic Alliteration Test corpus from ITU-T Recommendation P.807 \cite{itut_p807_2016}. The corpus comprises 96 minimal pairs (e.g., back / bag), designed to differ in exactly one phoneme and to probe six distinctive features: compactness, graveness, nasality, sibilation, sustention, and voicing \cite{itut_p807_2016}, in both word-initial and word-final positions.
 Clean speech data were collected through crowdsourcing by Lechler and Wojcicki \cite{lechler2024crowdsourced} from native English speakers. All recordings were manually checked. For each pair, we retain four talkers, two females and two males, with the talker set varying across pairs. To meet the carrier sentence constraint detailed in \ref{carrier_sentence}, we removed 2 stimuli. Stimuli span six conditions: (1) clean speech; (2–5) four anonymizers corresponding to open-source VPC baselines, B2 (McAdams) \cite{patino2021speaker}, B3 (STTTS) \cite{meyer2023prosody}, B4 (Neural Audio Codec) \cite{panariello2024speaker}, and B5 (ASR-BN) \cite{champion2022are}; and (6) a TTS condition (Qwen3-TTS-12Hz-1.7B-CustomVoice, with the English speaker "Ryan") \cite{hu2026qwen3tts}, in which audio is generated from the text only, rather than from the clean recording. In total, we gathered $2006$ audio stimuli. We then level-normalized them to -26 dB RMS to avoid volume-influenced bias.

\subsection{Crowdsourcing human evaluation}
\label{ssec:subhead2}

We developed a web-based interface to administer the listening test, in which the entire pipeline is automated: trap-question verification, participant payment, and rejections are handled without manual intervention. The interface is linked to the Prolific platform where participants are recruited.

\subsubsection{Two-stage design}
The study is split into two stages, a screening study and a rating study, with eight times fewer screening slots than rating slots, since each screened worker can complete up to eight rating jobs. Only participants who passed the screening were granted access to the rating study.

\subsubsection{Qualification job} Prolific's prescreening filters ensured that participants reported no hearing difficulties or cochlear implants, were native English speakers and were born and residing in the United Kingdom, United States of America, or Canada. Screening itself includes a headphone check using the Huggins pitch test \cite{milne2021online}. Participants were required to re-pass the screening every hour; we assume that within each one-hour window they remain in a quiet environment wearing headphones, as recommended by Naderi et al. \cite{naderi2020speech}.

\subsubsection{Rating job} Each rating job begins with one example question that allows participants to adjust their headphone volume, which they are then instructed not to change. The job consists of 27 questions presented in randomized order, with the six conditions equally represented, including three trap questions that all must be answered correctly. A rating job takes approximately 90 s, and each worker could complete up to eight jobs; 84 jobs in total were used to cover the 2,006 stimuli, with a target of eight ratings per stimulus. The minimal-pair word choices are displayed only after the first playback to avoid biasing the initial listening; participants may replay the stimulus as often as desired, and the number of replays is recorded.

Intelligibility is computed following \cite{itut_p807_2016}:
\begin{equation}
\label{eq:intel}
I_{human} = \frac{R - W}{R + W},
\end{equation}
with $R$ and $W$ as the number of right and wrong responses.

\subsection{ASR-based approximation}
\label{ssec:subhead3}

To approximate human judgments automatically, we let ASR models ``take the
test'' under the same two-alternative forced-choice protocol as the human
participants. We consider phoneme prediction models from the Hugging Face platform and models from \cite{srivastav2026open} that implement the prefix\_allowed\_tokens\_fn function. These comprise 2 non-autoregressive models trained via Connectionist Temporal Classification (CTC), and 10 autoregressive models.
Two quantities are recorded per stimulus: the correctness of the ASR's decision and its associated posterior probability.

\subsubsection{CTC-based model posterior probability} Given an input audio sequence $x$ and two candidate target words $y_A$ and
$y_B$, we compute their probability of the CTC sequence with the ASR model $m$ and its associated tokenizer $T$. The negative CTC losses serve as unnormalized log-scores:
\begin{equation}
s_i(x) = -\mathcal{L}_{\mathrm{CTC}}\bigl(x, T(y_i); m\bigr)
\qquad i \in \{A, B\},
\end{equation}
which we normalize with a softmax over the two candidates:
\begin{equation}
P(y_i \mid x, \mathcal{Y}) =
\frac{e^{s_i(x)}}{\displaystyle\sum_{y_j \in \mathcal{Y}} e^{s_j(x)}},
\qquad \mathcal{Y} = \{y_A, y_B\}.
\end{equation}

\subsubsection{Autoregressive model posterior probability} For autoregressive models, the two candidates are compared at the first token
position where they differ. Let the tokenized targets, of lengths $L_A$ and $L_B$ respectively, be
\begin{equation*}
\mathbf{y}_A = (a_1, \ldots, a_{L_A}), \qquad
\mathbf{y}_B = (b_1, \ldots, b_{L_B}),
\end{equation*}
and let $k$ denote the first differing position
$k = \min\bigl\{ t : a_t \neq b_t \bigr\}.$
The two targets share the same prefix
$\mathbf{y}_{<k} = (a_1, \ldots, a_{k-1}) = (b_1, \ldots, b_{k-1})$,
at which we extract the output probabilities of the next two competing tokens:
\begin{equation*}
p_A = p\bigl(a_k \mid x, \mathbf{y}_{<k}; M\bigr), \qquad
p_B = p\bigl(b_k \mid x, \mathbf{y}_{<k}; M\bigr).
\end{equation*}
The decoding is constrained to the
token sequences compatible with the two candidates, and the probabilities at
the divergence point yield a binary posterior:
\begin{equation*}
P(y_i \mid x, \{y_A, y_B\}) = \frac{p_i}{p_A + p_B},
\qquad i \in \{A, B\}.
\end{equation*}

\subsubsection{Carrier sentence}
\label{carrier_sentence}
Because ASR models may be sensitive to linguistic context, we compare
conditions with and without a \emph{carrier sentence}. To avoid biasing the ASR's response, the carrier contains no semantic cues related to the target words. As the original data contain only isolated words,
we construct carriers by concatenating five words from the same speaker; this
length was chosen so that virtually all stimuli from the original corpus could
be retained (some talkers have too few recordings to build longer carriers).
The carrier text is provided to the ASR as ground truth, so ASR
errors do not affect the comparison.

\subsubsection{Ensembling} From the posterior probabilities of the ASR models, we derive three intelligibility
estimates. 
\begin{itemize}[leftmargin=*, labelwidth=0pt, labelsep=0.5em, itemindent=0pt, align=left]
\item \textbf{Hard voting} mirrors the human formula~\eqref{eq:intel}:
\begin{equation}
I_{\mathrm{hard}} = \frac{R - W}{R + W},
\end{equation}
where $R$ and $W$ are the numbers of right and wrong model decisions.
\item \textbf{Soft voting} weights each decision by its posterior:
\begin{equation}
I_{\mathrm{soft}} = \frac{\sum_{k=1}^{K}\sum_{m=1}^{M} \left( P_{r,k}^{(m)} - P_{w,k}^{(m)} \right)}{\sum_{k=1}^{K}\sum_{m=1}^{M} \left( P_{r,k}^{(m)} + P_{w,k}^{(m)} \right)}.
\end{equation}
where $K$ is the number of stimuli at the aggregation level, $k$ is the stimulus among $K$ and $M$ is the number of models. 

\item Assuming conditional independence between the model outputs given the
input, the joint evidence for the correct word is the product of the posteriors per model.
We thus define the \textbf{log-likelihood ratio (LLR)}:
\begin{equation}
I_{\mathrm{LLR}} = \frac{1}{K}\sum_{k=1}^{K}
\frac{1}{M}\sum_{m=1}^{M}
\log\frac{P_{r,k}^{(m)}}{P_{w,k}^{(m)}}.
\end{equation}
\end{itemize}
We then fit a linear mapping $I_{human} = aI + b$ between ASR-based and human intelligibility scores, reporting the coefficient of determination $r^2$.

\section{Experiments and Results}
\label{sec:typestyle}

\subsection{Human intelligibility ratings}
\label{ssec:subhead4}

\begin{table}[t]
\centering
\caption{Intelligibility (human ratings) and reported WER by the VPC utility evaluation protocol, per condition. Note: Since Qwen-TTS is not an anonymizer, the WER from the VPC evaluation does not apply.}
\label{tab:results}
\begin{tabular}{lcc}
\toprule
\textbf{Condition} & \textbf{Intelligibility (\%) $\uparrow$} & \textbf{WER (\%) $\downarrow$} \\
\midrule
B2 (McAdams)      & 72.2 & 9.96 \\
B3 (STTTS)       & 70.4 & 4.31 \\
B4 (Neural Codec) & 85.2 & 5.90 \\
B5 (ASR-BN)        & 78.2 & 4.44 \\
Qwen-TTS          & 90.2 & N/A  \\
Clean (reference) & 91.8 & 1.84 \\
\bottomrule
\end{tabular}
\end{table}

A total of 188 participants took the test, at a cost of 371€. Of these, 127 passed the screening and 102 completed at least one rating job. Three failed the trap questions; their records were excluded from all subsequent analyses.
In total, 677 rating jobs were recorded, yielding 16,168 individual ratings with an average of 6.8 jobs per participant. The mean completion time was 97 s per rating job, and a mean response time of 3 s per stimulus.
The intelligibility results are reported in Table \ref{tab:results} and compared to the objective metric used by the Voice Privacy Challenge.
Human intelligibility and the WER reported by the VPC disagree in both ranking and magnitude, indicating that phoneme-level intelligibility and WER computed on the LibriSpeech test set are not interchangeable.

\subsection{Human-ASR correlation and approximation}
\label{ssec:subhead5}

\begin{table}[t]
\centering
\caption{$r^2$ between human
intelligibility \& ASR-based scores ($I_{\mathrm{soft}}$, $I_{\mathrm{hard}}$,
$I_{\mathrm{LLR}}$), carrier condition (without / with / both), and aggregation level ($N$ = number of data points). A Steiger's $Z$ test confirms that $r_{I_{\mathrm{soft}}}$ and $r_{I_{\mathrm{hard}}}$
differ significantly at the stimulus level ($p<0.05$), but we cannot reject the null hypothesis at higher aggregation levels. $r_{w/o}$ and $r_{w/ + w/o}$ differ significantly at the stimulus level ($p<0.05$).}
\label{tab:r2}
\begin{tabular}{llccc}
\toprule
\textbf{Aggregation} & \textbf{Carrier} & ${I_{\mathrm{soft}}}\uparrow$ & ${I_{\mathrm{hard}}}\uparrow$ & ${I_{\mathrm{LLR}}}\uparrow$ \\
\midrule
\multirow{3}{*}{\shortstack{stimuli\\($N = 2006$)}}
 & w/o        & 0.4039 & 0.4100 & 0.3164 \\
 & w/         & 0.3371 & 0.3484 & 0.3224 \\
 & w/ + w/o   & \textbf{0.4424} & \textbf{0.4513} & \textbf{0.3852} \\
\midrule
\multirow{3}{*}{\shortstack{$C$ \\($N = 36$)}}
 & w/o        & \textbf{0.8159} & 0\textbf{.8208} & 0.7461 \\
 & w/         & 0.7284 & 0.7251 & 0.7198 \\
 & w/ + w/o   & 0.8085 & 0.8079 & \textbf{0.7708} \\
\midrule
\multirow{3}{*}{\shortstack{$B$ \\($N = 12$)}}
 & w/o        & \textbf{0.9648} & 0.9651 & 0.9323 \\
 & w/         & 0.8856 & 0.8914 & 0.8669 \\
 & w/ + w/o   & 0.9611 & \textbf{0.9657} & \textbf{0.9381} \\
\midrule
\multirow{3}{*}{\shortstack{$A$ \\($N = 6$)}}
 & w/o        & \textbf{0.9850} & \textbf{0.9845} & \textbf{0.9482} \\
 & w/         & 0.8777 & 0.8836 & 0.8709 \\
 & w/ + w/o   & 0.9685 & 0.9714 & 0.9453 \\
\bottomrule
\end{tabular}
\end{table}

We report intelligibility at three levels of granularity as proposed by \cite{itut_p807_2016}: $(A)$ a single global value per condition, $(B)$  two per-test-type values (word-initial and word-final phoneme-variation) and $(C)$ six per-feature values (compactness, graveness, nasality, sibilation, sustention, voicing).

Table~\ref{tab:r2} shows the coefficient of determination $r^2$ at
each aggregation level, comparing the three proposed metrics to the human intelligibility. Across all aggregations, $I_{\mathrm{soft}}$ and $I_{\mathrm{hard}}$ perform
almost identically and both consistently outperform $I_{\mathrm{LLR}}$. The carrier sentence also affects the scores, but its impact depends on the
aggregation level: at the stimulus level ($N = 2006$), combining conditions
with and without the carrier significantly improves the fit ($r^2$ from $0.40$ to $0.44$
for $I_{\mathrm{soft}}$, $p<0.05$ on a Steiger's $Z$ test), whereas at coarser levels (e.g. $N = 6$), this
combination seems detrimental as the condition without carrier alone performs best. This suggests that the carrier helps the ASR at the individual-stimulus level but introduces noise once the scores are aggregated.

\begin{table}[bh]
\centering
\caption{$r^2$ between listening-test intelligibility according to Eq.~\eqref{eq:intel} and
$I_{\mathrm{soft}}$, per ASR model and carrier condition at the stimulus aggregation level. Model names are abbreviated. The last row reports the ensemble; the fourth column gives the score when stimuli with and
without carrier are combined.}
\label{tab:r2_models}
\begin{tabular}{lccc}
\toprule
\textbf{Model} & \textbf{w/o} $\uparrow$ & \textbf{w/} $\uparrow$ & \textbf{w/ + w/o} $\uparrow$ \\
\midrule
Cohere Transcribe          & 0.2392 & 0.1922 & 0.2874 \\
wav2vec2-lv-60-espeak      & 0.1739 & 0.2163 & 0.2419 \\
wav2vec2-xlsr-53-espeak    & 0.1714 & 0.1440 & 0.2041 \\
Whisper-large-v3           & 0.1379 & 0.2069 & 0.2690 \\
Whisper-large-v3-turbo     & 0.1185 & 0.1849 & 0.2424 \\
Distil-Whisper-large-v3.5  & 0.1888 & 0.1943 & 0.2826 \\
Canary-Qwen-2.5B           & 0.1840 & 0.1451 & 0.2435 \\
Qwen3-ASR-1.7B             & 0.1679 & 0.1852 & 0.2477 \\
Qwen3-ASR-0.6B             & 0.1658 & 0.1743 & 0.2664 \\
Granite-4.0-1B-Speech      & 0.1732 & 0.1950 & 0.2616 \\
Granite-Speech-4.1-2B      & 0.1642 & 0.1742 & 0.2415 \\
Granite-Speech-4.1-2B+     & 0.1667 & 0.1743 & 0.2507 \\
\midrule
\textbf{All models (ensemble)} & \textbf{0.4039} & \textbf{0.3371} & \textbf{0.4424} \\
\bottomrule
\end{tabular}
\end{table}

\begin{figure}[t]
\centering
\includegraphics[width=\columnwidth]{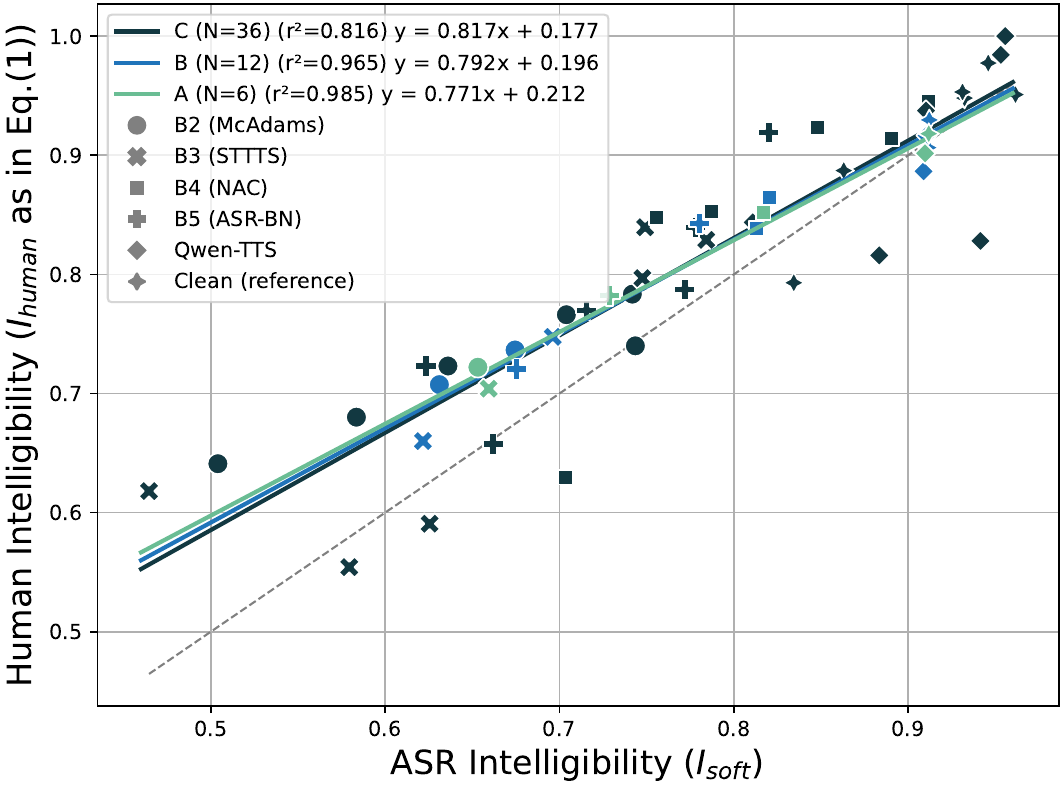}
\caption{Fitted linear regression $I_{human} = a\,I_{\mathrm{soft}} + b$ per aggregation level (A: green, B: blue, C: black) w/o carrier sentence. For reference, the dotted gray line shows identity. The null hypotheses of homoscedasticity (Breusch--Pagan test) and normally
distributed residuals (Shapiro--Wilk test) cannot be rejected ($p>0.05$),
in agreement with the quantile-quantile-plots, except at level B where normality is
rejected.}
\label{fig:regline}
\end{figure}

Reported in Table \ref{tab:r2_models}, at the stimulus level, no single ASR model dominates across
both carrier conditions. This carrier-dependent ranking confirms that models respond heterogeneously
to acoustic context. Combining the posteriors across all models yields
$r^2 = 0.40$ without and $0.34$ with the carrier, a 69\% relative improvement over the best model, and combining both conditions further raises the
ensemble $r^2$ to $0.44$. We suggest that ensembling thus acts as a robustness mechanism, compensating for the sensitivities of individual ASR systems. Furthermore, a Spearman-Brown corrected reliability \cite{spearman1910correlation, brown1910experimental} yields $r_{xx'} = 0.81$ mean over 100 random half-splits on human ratings, indicating that our best metric captures $0.45/0.81=56\%$ of the reliably measurable variance at the stimulus level. Finally, Fig.~\ref{fig:regline} shows the linear mapping between ASR-based and human intelligibility, indicating a highly predictive $r^2 = 0.985$ for condition-wise aggregation.

\subsection{Discussion and limitations}
\label{ssec:subhead6}

Crowdsourced ratings are inherently noisier than laboratory data, as we rely on self-reported English fluency, and as the participants may be using diverse setups.
The $I_{\mathrm{LLR}}$ metric
assumes independent model outputs---a typical assumption, but not true for our data. We performed a calibration analysis
by computing the Expected Calibration Error, which led to the observation that some models are not well-calibrated, possibly explaining $I_{\mathrm{LLR}}$'s weaker performance. 
A leave-one-out 5 fold cross validation confirmed that some models contribute more than others. Given the limited data, we still
retained a univariate mapping with all models to avoid overfitting. 
Due to the typically high acoustic quality of the considered anonymization systems, our corpus covers only six conditions in
the upper range of the intelligibility scale; the regression on
strongly degraded, out-of-distribution speech remains to be studied, e.g., for applications where noisy or otherwise difficult speech signals are to be anonymized.

\section{Conclusion}
\label{sec:majhead}

We proposed the first phoneme-level intelligibility evaluation of speech
anonymizers.
Our results show that simple hard-voting ASR metrics strongly correlate with human ratings, provided that multiple models are ensembled. Combining stimuli with and without a carrier sentence further improves
the correlation at the stimulus level, while confidence-based metrics bring no gain over correctness-based
decisions, likely due to insufficient model calibration.

\clearpage
\vfill\pagebreak
\section{Compliance with Ethical Standards}
\label{sec:ethic}
This study was approved by the Ethics Committee of the Technische Universität Berlin; all participants provided informed consent, and only anonymized data are released.

\section{Acknowledgments}
\label{sec:ack}

This work was funded by the European Union’s Horizon Europe research and innovation programme grant No 101168193.

\bibliographystyle{IEEEbib}
\bibliography{strings,refs}

\end{document}